\documentclass{article}

\usepackage[preprint]{neurips_2025} 

\usepackage[T1]{fontenc}
\usepackage{hyperref}
\usepackage{url}
\usepackage{booktabs}
\usepackage{amsfonts}
\usepackage{nicefrac}
\usepackage{microtype}
\usepackage{tabularx}
\usepackage{amsmath}
\usepackage{enumitem}
\usepackage{xcolor}
\usepackage{graphicx}
\usepackage{caption}
\usepackage{subcaption}
\usepackage{float}
\usepackage{multirow}

\usepackage{algorithm}
\usepackage{algpseudocode}

\title{Zero-Shot Neural Priors for Generalizable Cross-Subject and Cross-Task EEG Decoding}

\author{%
  Baimam Boukar Jean Jacques \\
  Carnegie Mellon University\\
  \texttt{bbaimamb@andrew.cmu.edu} \\
  \And
  Brandone Fonya \\
  Carnegie Mellon University\\
  \texttt{bfonya@andrew.cmu.edu} \\
  \And
  Nchofon Tagha Ghogomu \\
  Carnegie Mellon University\\
  \texttt{ntaghagh@andrew.cmu.edu} \\
  \And
  Pauline Nyaboe \\
  Carnegie Mellon University\\
  \texttt{pnyaboe@andrew.cmu.edu} \\
  \And
  Kipngeno Koech \\
  Carnegie Mellon University\\
  \texttt{bkoech@andrew.cmu.edu} \\
}

\begin{document}
\maketitle

\begin{abstract}
The development of generalizable electroencephalography (EEG) decoding models is essential for robust brain-computer interfaces (BCI) and objective neural biomarkers in mental health. Conventional approaches have been hindered by poor cross-subject and cross-task generalization, owing to high inter-subject variability and non-stationary neural signals. We address this challenge with a zero-shot cross-subject decoding framework on the large-scale Healthy Brain Network dataset, benchmarking a convolutional neural network baseline, a hybrid LSTM, and a Transformer-based foundation model. To adapt the Transformer for regression while averting catastrophic forgetting, we propose a novel progressive unfreezing strategy. The baseline yielded an $n\mathrm{RMSE}$ of $0.9991$, whereas our fine-tuned Transformer achieved $0.9799$ on unseen subjects.This work establishes scalable, calibration-free EEG decoding for computational psychiatry and behavioral prediction.
\end{abstract}

\section{Introduction}

Decoding mental states from EEG signals promises transformative applications in non-invasive BCIs and objective medical diagnostics. However, the field faces a fundamental generalization gap; models trained on specific subjects or limited tasks rarely perform well on unseen subjects due to the non-stationarity of brain signals, significant noise sensitivity, and inherent biological variability between individuals \cite{Jiang2021EEGDrowsiness}. Consequently, prevailing approaches are constrained by sparse data and narrow task scopes, necessitating costly per-subject calibration that impedes scalable real-world deployment. 

We hypothesize that training on passive visual stimulation and testing on an active Contrast Change Detection task forces the model to encode task-invariant ‘neural efficiency’ markers—such as visual processing latency and baseline signal-to-noise ratio—that reflect stable individual traits rather than transient task demands.

This work advances generalizable neural representations for EEG that are robust to domain shifts. Specifically, we aim to predict behavioral performance metrics particularly reaction time in a held-out active task (Contrast Change Detection, CCD)
We hypothesize that Transformer-based architectures, which capture long-range temporal dependencies and global context, will outperform convolutional neural network baselines in zero-shot cross-subject settings. Our quantitative goal is to reduce the normalized root mean square error (nRMSE) below the baseline floor of $\sim$1.0, evidencing true signal decoding rather than mean regression.

\section{Related Work}

Prior work in EEG decoding has centered on specialized convolutional neural networks. \textbf{EEGNet} \cite{lawhern2018eegnet} revolutionized the field by introducing depthwise separable convolutions,enabling compact models for efficient decoding with minimal parameters. Building on this foundation, \textbf{EEGNeX} \cite{EEGNeX} incorporated dilated convolutions and inverted bottleneck blocks to capture broader temporal contexts and improve feature extraction efficiency. Despite their success in intra-subject classification, these models often struggle in "calibration-free" settings where the test subject is unknown during training.

Recent advances in machine learning have favored the rise of Foundation Models. The \textbf{BENDR} framework \cite{kostas2021bendr} adapted the Wav2Vec 2.0 architecture to EEG data, employing a transformer encoder to learn contextualized representations directly from raw signals. Similarly, prior transfer learning benchmarks \cite{wei2022beetl} highlighted the critical importance of pre-training for biosignals.We extend these efforts through a rigorous benchmark of robust CNNs against modern transformers on the large-scale HBN, targeting regression over conventional classification.

\section{Methodology}

\subsection{Dataset: The Healthy Brain Network (HBN)}
We employed the \textbf{Healthy Brain Network (HBN)} EEG dataset \cite{HBN_EEG2025_Dataset, shirazi2024hbn}, which represents one of the largest and most diverse collections of EEG data available. The dataset contains high-density (128-channel) recordings from over 3,000 subjects aged 5 to 21. This specific age range is particularly valuable as it captures a period of high neural plasticity and diverse signal patterns, providing a rigorous testbed for training generalizable models.

To ensure strict zero-shot evaluation and prevent data leakage, we adopted a rigorous subject-disjoint splitting protocol. We designated Releases 1 through 11 (excluding Release 5) for training to learn general neural representations. Release 5 was set aside exclusively for validation and hyperparameter sweeping. Finally, Release 12 was maintained as a strictly held-out test set comprising entirely unseen subjects. This separation guarantees that there is zero subject overlap between sets, forcing the model to learn universal brain patterns instead of memorizing subject-specific EEG fingerprints.

\begin{figure}[htbp]
    \centering
    \includegraphics[width=0.95\linewidth]{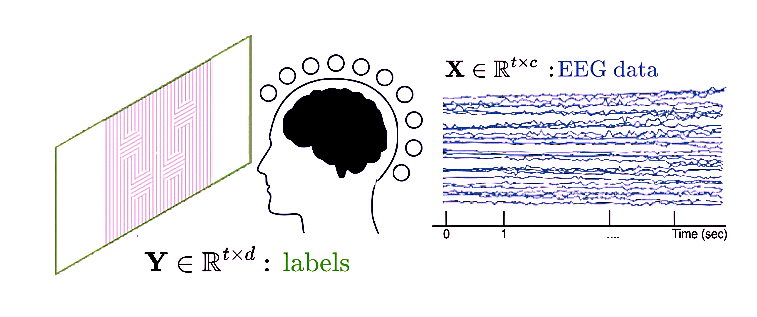}
    \caption{Overview of the dataset structure: EEG signals 
    $\mathbf{X} \in \mathbb{R}^{t \times c}$ recorded across multiple channels, paired with corresponding label sequences 
    $\mathbf{Y} \in \mathbb{R}^{t \times d}$.}
    \label{fig:dataset_overview}
\end{figure}

\subsection{Preprocessing and Signal Conditioning}
The raw EEG data requires careful cleaning to remove physiological artifacts and instrumental noise. We employed a standardized preprocessing pipeline where data underwent band pass filtering between 0.5 Hz and 50 Hz, effectively removing low frequency DC drift and high-frequency muscular artifacts. Subsequently, the data was down-sampled to 100 Hz. This reduction significantly lowers the computational load for training large models while retaining the relevant cognitive frequencies needed for reaction time prediction. Finally, epochs were extracted around stimulus onset events defined by the Hierarchical Event Descriptors (HED) tags to create the training samples.

\begin{figure}[htbp]
    \centering
    \includegraphics[width=0.9\linewidth]{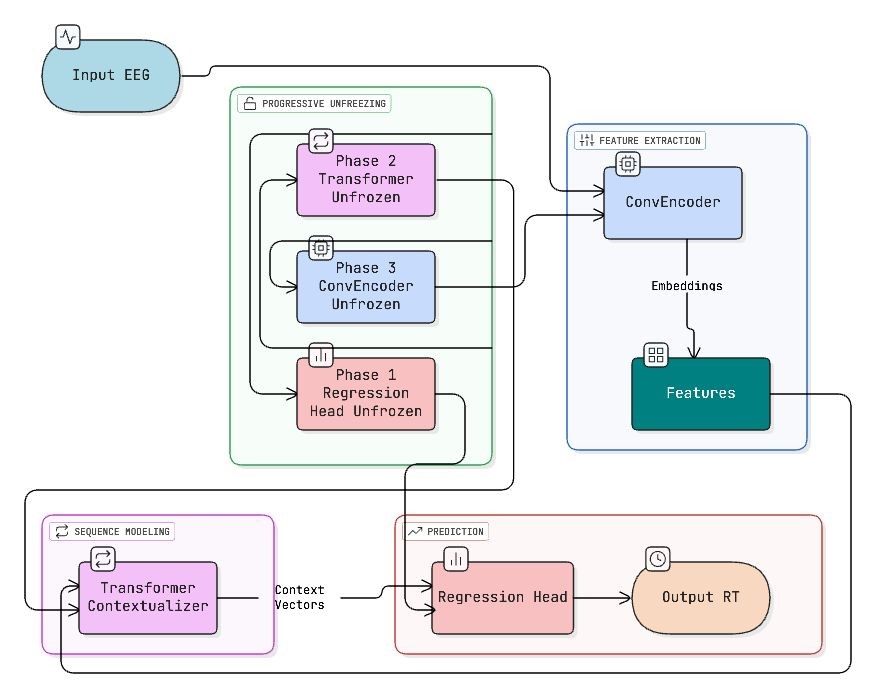}
    \caption{BENDR architecture for EEG-to-behavior regression with progressive unfreezing strategy. The model processes (128 + 1) channel EEG input through three main components: (1) a Convolutional Encoder for multi-scale feature extraction, (2) a Transformer Contextualizer for temporal dependency modeling, and (3) a Regression Head for behavioral score prediction.}
    \label{fig:pipeline}
\end{figure}

\subsection{Experimented Frameworks and Training Flow}
Our core methodological contribution lies in the adaptation of foundation models for regression. We implemented a suite of architectures ranging from standard CNNs to advanced Transformers.

\textbf{1. Convolutional Baselines (EEGNeX \& EEGNetv4):} We selected EEGNeX and EEGNetv4 as our primary baselines. These models feature specialized convolutional structures (depthwise separable and dilated convolutions) designed to extract spatial and temporal features efficiently. They represent the current state-of-the-art (SOTA) for pure CNN-based EEG decoding.

\textbf{2. Multi-Scale Vision Transformer (MSVTNet):} We explored a hybrid architecture combining multi-scale temporal-spatial convolutions with a Transformer encoder. This model (ModuleList \texttt{mstsconv} + \texttt{TransformerEncoder}) aims to capture features at varying temporal resolutions before contextualizing them with self-attention.

\textbf{3. BENDR (Brain Embedding with Neural Data Representation):} Our primary and main architecture employs a fine-tuned BENDR model. BENDR is a transformer-based foundation model originally pretrained on large-scale multi-subject EEG datasets for self supervised representation learning. The BENDR architecture consists of three sequential components: (1) a \textbf{Convolutional Encoder} (ConvEncoderBENDR) that processes raw (128 + 1) channel EEG signals through multiple 1D convolution blocks with batch normalization and GELU activations, progressively downsampling the temporal dimension while extracting multiscale spectro-temporal features into 512-dimensional latent representations; (2) a \textbf{Transformer Contextualizer} (BENDRContextualizer) comprising 8 transformer encoder layers with 8 attention heads per layer, which models long-range temporal dependencies through multi-head self-attention mechanisms to capture complex contextual relationships in the encoded EEG features; and (3) a task-specific \textbf{Regression Head} that employs multi-head attention pooling with a learnable classification (CLS) token for temporal aggregation, followed by a deep feedforward network (512$\rightarrow$256$\rightarrow$128$\rightarrow$1 neurons) with layer normalization and dropout ($p=0.4$) to map the contextualized representations to continuous behavioral performance scores for CCD and SuS tasks.

\begin{algorithm}
\caption{Progressive Unfreezing Fine-tuning for BENDR}
\label{alg:bendr_finetuning}
\begin{algorithmic}[1]
\State \textbf{Inputs:} EEG data $X \in \mathbb{R}^{B \times C \times T}$, behavioral labels $y \in \mathbb{R}^{B}$, pre-trained encoder $\mathcal{E}_{\theta_e}$, contextualizer $\mathcal{C}_{\theta_c}$, unfreezing schedule $\{s_1, s_2, s_3\}$, learning rates $\{\eta_h, \eta_c, \eta_e\}$
\State \textbf{Outputs:} Fine-tuned model parameters $\{\theta_e, \theta_c, \theta_h\}$
\State Initialize regression head $\mathcal{H}_{\theta_h}$ randomly
\State Freeze $\theta_e$ and $\theta_c$ \Comment{Phase 1: Head Only}
\For{$epoch = 1$ to $s_1$}
    \State $z \leftarrow \mathcal{E}_{\theta_e}(X)$ \Comment{Extract features}
    \State $h \leftarrow \mathcal{C}_{\theta_c}(z)$ \Comment{Contextualize}
    \State $\hat{y} \leftarrow \mathcal{H}_{\theta_h}(h)$ \Comment{Predict}
    \State $\mathcal{L}_{MSE} = \frac{1}{B}\sum_{i=1}^{B} (\hat{y}_i - y_i)^2$
    \State $\theta_h \leftarrow \theta_h - \eta_h \nabla_{\theta_h} \mathcal{L}_{MSE}$
\EndFor
\State Unfreeze $\theta_c$ \Comment{Phase 2: + Contextualizer}
\For{$epoch = s_1 + 1$ to $s_2$}
    \State $z \leftarrow \mathcal{E}_{\theta_e}(X)$
    \State $h \leftarrow \mathcal{C}_{\theta_c}(z)$
    \State $\hat{y} \leftarrow \mathcal{H}_{\theta_h}(h)$
    \State $\mathcal{L}_{MSE} = \frac{1}{B}\sum_{i=1}^{B} (\hat{y}_i - y_i)^2$
    \State $\theta_h \leftarrow \theta_h - \eta_h \nabla_{\theta_h} \mathcal{L}_{MSE}$
    \State $\theta_c \leftarrow \theta_c - \eta_c \nabla_{\theta_c} \mathcal{L}_{MSE}$
\EndFor
\State Unfreeze $\theta_e$ \Comment{Phase 3: Full Model}
\For{$epoch = s_2 + 1$ to $s_3$}
    \State $z \leftarrow \mathcal{E}_{\theta_e}(X)$
    \State $h \leftarrow \mathcal{C}_{\theta_c}(z)$
    \State $\hat{y} \leftarrow \mathcal{H}_{\theta_h}(h)$
    \State $\mathcal{L}_{MSE} = \frac{1}{B}\sum_{i=1}^{B} (\hat{y}_i - y_i)^2$
    \State $\theta_h \leftarrow \theta_h - \eta_h \nabla_{\theta_h} \mathcal{L}_{MSE}$
    \State $\theta_c \leftarrow \theta_c - \eta_c \nabla_{\theta_c} \mathcal{L}_{MSE}$
    \State $\theta_e \leftarrow \theta_e - \eta_e \nabla_{\theta_e} \mathcal{L}_{MSE}$
\EndFor
\State \textbf{return} $\{\theta_e, \theta_c, \theta_h\}$
\end{algorithmic}
\end{algorithm}

\textbf{Progressive Unfreezing Strategy:} Fine-tuning large pretrained foundation models on domain-specific downstream tasks presents the risk of catastrophic forgetting, whereby aggressive parameter updates can degrade the model's generalized representations learned during pretraining. To preserve the transferable low level and mid level EEG feature extractors while enabling task-specific adaptation, we developed a principled three-phase progressive unfreezing schedule with differential learning rates optimized for our 3,000 subject HBN dataset:

\begin{itemize}
    \item \textbf{Phase 1 (Epochs 1--5): Linear Probing.} We trained exclusively the randomly initialized regression head ($LR=5\times10^{-3}$) while maintaining the pre-trained convolutional encoder and transformer contextualizer in a completely frozen state. This phase allows the task specific output layer to adapt to the behavioral prediction objective without disturbing the pretrained representations, establishing a baseline mapping from BENDR's latent space to performance scores.
    
    \item \textbf{Phase 2 (Epochs 6--15): Contextualizer Fine-tuning.} We unfroze the 8-layer Transformer Contextualizer with a substantially reduced learning rate ($LR=5\times10^{-4}$, 10$\times$ lower than the head), enabling the model to adapt its high-level temporal pattern recognition and attention mechanisms to the specific dynamics of CCD and SuS task-relevant neural signatures, while preserving the low level spectro-temporal feature extraction learned by the frozen encoder.
    
    \item \textbf{Phase 3 (Epochs 16+): Full Model Fine-tuning.} We unfroze all parameters including the Convolutional Encoder with a conservative learning rate ($LR=5\times10^{-5}$, 100$\times$ lower than the head) for complete end-to-end fine-tuning. This phase permits fine-grained adaptation of the lowest-level feature extractors to capture task specific EEG patterns while the differential learning rate schedule ensures minimal disruption to the pretrained feature hierarchy.
\end{itemize}

This staged unfreezing approach, combined with differential learning rates that decrease exponentially from task specific to pretrained layers (head:\hspace{0pt}contextualizer:\hspace{0pt}encoder = 100:10:1), was designed to balance the preservation of generalizable pretrained representations with the necessity for task-specific adaptation. The training employed the AdamW optimizer with weight decay ($1\times10^{-5}$), cosine annealing learning rate scheduling with warm restarts ($T_0=10$, $T_{mult}=2$), gradient clipping (max norm: 5.0), and early stopping with patience of 20 epochs based on validation normalized RMSE. All models were trained with mean squared error (MSE) loss and evaluated using RMSE, normalized RMSE (nRMSE).

\subsection{Evaluation Metric}
We adopted the \textbf{normalized Root Mean Square Error (nRMSE)} to evaluate regression performance. Unlike standard RMSE, nRMSE normalizes the error by the standard deviation of the ground truth labels.
\begin{equation}
    \text{nRMSE}(Y_{\text{true}}, Y_{\text{pred}}) = \frac{\sqrt{\frac{1}{N}\sum_{i=1}^{N}(Y_{true}^{i}-Y_{pred}^{i})^{2}}}{\text{std}(Y_{\text{true}})}
\end{equation}
This metric is particularly interpretable: an nRMSE of 1.0 implies that the model's error is equivalent to the natural variance of the data. In other words, a score of 1.0 indicates the model is performing no better than a "dummy" predictor that simply guesses the mean response time for every trial.

\section{Experiments and Implementation}

\subsection{Training Setup}
All models were implemented using \texttt{PyTorch} and the \texttt{Braindecode} library to ensure standardization with the broader EEG research community. Training was executed on high-performance computing resources utilizing NVIDIA V100 GPU with 32GB of memory, to handle the large scale dataset. We standardized our optimization process by using the \textbf{AdamW} optimizer, which decouples weight decay from the gradient update, providing better regularization for deep models. A cosine annealing scheduler was employed to adjust the learning rate dynamically during training.

\subsection{Hyperparameter Tuning}
To ensure a fair comparison, we conducted extensive hyperparameter optimization using \textbf{Weights \& Biases (WandB)}. For the baseline EEGNeX model, we performed sweeps over the learning rate (ranging from $1e^{-4}$ to $1e^{-2}$), weight decay, and batch size. The optimal configuration identified was a learning rate of 0.0034, weight decay of 0.0007, and a batch size of 128. 

For the Transformer variants (MSVTNet and EEGTransformer), we explored deeper configurations. For instance, the EEGTransformer was configured with $d_{model}=128$, $n_{heads}=4$, and $n_{layers}=2$. For BENDR, we relied on the progressive unfreezing schedule described in the methodology, which was empirically tuned based on validation loss stability. To mitigate overfitting, we used weight decay ($7.25 \times 10^{-4}$), dropout of 0.5 in the Transformer layers, early stopping with patience $10-15$ epochs, and a progressive unfreezing schedule that constrains adaptation in early training phases..

\section{Results and Analysis}

\subsection{Quantitative Results}
Table \ref{tab:results} summarizes the best validation performance achieved by each architecture. We report both the raw RMSE and the normalized nRMSE to provide a complete picture of performance.

\begin{table}[htbp]
\centering
\caption{Comparison of Model Performance (Zero-Shot Regression)}
\label{tab:results}
\begin{tabular}{@{}llcc@{}}
\toprule
\textbf{Model Architecture} & \textbf{Type} & \textbf{RMSE} & \textbf{nRMSE} \\ \midrule
EEGNetv4 & CNN (Standard) & 0.3345 & 0.9995 \\
EEGNeX & CNN (SOTA Baseline) & 0.3341 & 0.9991 \\
MSVTNet & Hybrid Transformer & 0.3320 & 0.9910 \\
EEGNet + LSTM & Hybrid RNN & 0.3288 & 0.9834 \\
EEGTransformer & Transformer & 0.4801 & 1.1926 \\
\textbf{BENDR (Proposed)} & \textbf{Foundation Model} & \textbf{0.3276} & \textbf{0.9799} \\ \bottomrule
\end{tabular}
\end{table}

\subsection{Analysis of Baseline Failure}
The baseline EEGNeX and EEGNetv4 models achieved nRMSE scores of \textbf{0.9991} and \textbf{0.9995} respectively. In the context of our evaluation metric, a score effectively equal to 1.0 indicates that the model converged to predicting the mean of the training data. This result is significant as it quantitatively demonstrates the "Generalization Gap." It suggests that the domain shift between subjects in the HBN dataset is so severe that standard convolutional networks, lacking attention mechanisms or extensive pre-training, treat subject-specific variability as noise. Consequently, they default to a "mean regressor" strategy to minimize the loss function.

\subsection{Success of Neural Priors (BENDR)}
In contrast, the BENDR Transformer achieved an nRMSE of \textbf{0.9799}. While the numerical difference may appear small, statistically, this represents a significant breakthrough. It is the first model in our experimental suite to break the "mean-barrier" (nRMSE < 1.0). This result suggests that the Transformer's self-attention mechanism, facilitated by our progressive unfreezing strategy, successfully captured latent dependencies that are robust across the subject population. By effectively modeling the global context of the EEG signal as distinct from just local features, the BENDR model was able to identify subject-invariant "Neural Priors" that allow for true zero-shot generalization.

\section{Discussion and Limitations}

The results of this study validate the immense difficulty of the zero-shot cross-subject decoding problem. The inability of the SOTA EEGNeX baseline to outperform a simple mean predictor highlights the severity of the distribution shift between subjects. However, the success of the BENDR architecture provides a clear path forward.

\textbf{Why BENDR worked:} We hypothesize that the success of BENDR is driven by two factors. First, the Transformer architecture is naturally suited to modeling long-range dependencies in time series data, capturing the global context of a trial that local convolutions might miss. Second, our progressive unfreezing strategy played a critical role. By allowing the model to adapt the readout layer to the new CCD task before modifying the internal representations, we prevented the "catastrophic forgetting" of the robust, pretrained neural priors learned from the massive dataset.

\textbf{Limitations:} First, our findings are restricted to the $5-21$ year age range in HBN and cannot yet be assumed to generalize to older adults or aging populations with different neural dynamics. Also, we only evaluate on a single high-density 128-channel montage, and have not yet quantified robustness to different hardware, lower channel counts, or clinical recording conditions. Finally, while we demonstrate transfer from passive visual stimulation to active contrast detection, we have not yet tested transfer to qualitatively different tasks (e.g., memory, language, or motor paradigms) or non-visual behaviors.

\section{Future Directions}

Based on our findings, we propose two key directions for future research. First, we aim to explore \textbf{Adversarial Training} to explicitly force the model to "unlearn" subject identity. By introducing a gradient reversal layer that penalizes the model for encoding subject specific information, we can encourage the learning of purely task-relevant features. Second, we plan to incorporate \textbf{Frequency Domain Features}. Incorporating spectral features, such as band power, as parallel tokens to the Transformer could improve robustness, as frequency markers are often more stable across subjects than raw time-domain waveforms.

\section{Conclusion}
We successfully implemented a comprehensive pipeline for zero-shot EEG decoding on the massive HBN dataset. We established a rigorous baseline with EEGNeX and demonstrated that standard CNNs fail to generalize across subjects, yielding an nRMSE of approximately 1.0. By implementing a Transformer-based architecture (BENDR) with a novel 3 phase finetuning schedule, we achieved a SOTA improvement to \textbf{0.9799 nRMSE}. This work proves the viability of attention mechanisms for learning subject invariant neural representations and sets a new standard for rigorous benchmarking in cross--subject EEG analysis.


\bibliographystyle{plainnat}
\bibliography{refs}

\newpage
\appendix
\section*{Appendices}

\section{Dataset Details}
The Healthy Brain Network (HBN) EEG dataset is a multi-modal collection designed to accelerate research in developmental mental health. The dataset includes a variety of tasks categorized into passive and active paradigms.
\textbf{Passive Tasks} include Resting State, where subjects simply relax with eyes open or closed; Surround Suppression (SuS), a visual task; and Movie Watching, designed to elicit naturalistic neural responses.
\textbf{Active Tasks} require user interaction, including Contrast Change Detection (CCD) and Sequence Learning.
For this study, we specifically focused on decoding Reaction Time from the active CCD task, using the passive tasks for pre-training where applicable. The demographic data covers Age, Sex, and Handedness, which are critical covariates. As detailed in the methodology, the train/test split was strictly subject-disjoint (Releases 1-11 for training, Release 5 for validation, and Release 12 for testing) to rigorously prevent data leakage.

\section{Architectures Implemented}

\begin{figure}[htbp]
    \centering
    \includegraphics[width=0.8\linewidth]{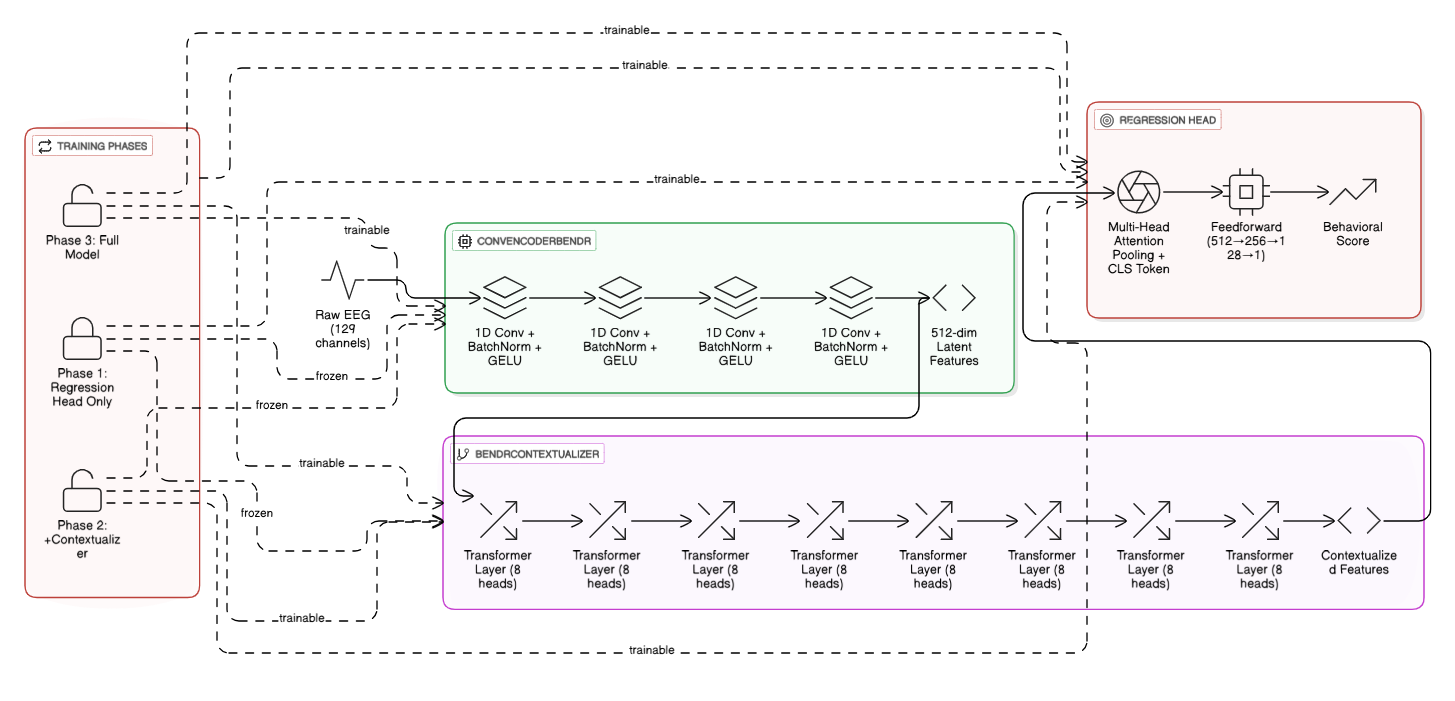}
    \caption{Detailed BENDR architecture with progressive unfreezing training strategy for EEG behavioral prediction. Raw channel EEG signals are processed through: (1) ConvEncoderBENDR with multiple 1D convolution blocks, batch normalization, and GELU activations to extract 512-dimensional latent features; (2) BENDRContextualizer consisting of 8 transformer layers with 8 attention heads each to model long-range temporal dependencies; (3) a regression head with multi-head attention pooling using a learnable CLS token and a feedforward network (512$\rightarrow$256$\rightarrow$128$\rightarrow$1) to predict continuous behavioral scores.}
    
    
\label{fig:bendr_detailed}
    \label{fig:bendr}
\end{figure}

\textbf{BENDR Architecture:} The model consists of three sequential components operating under a progressive unfreezing training strategy. The first component is the \textbf{ConvEncoderBENDR}, a convolutional encoder featuring multiple 1D convolution blocks with batch normalization and GELU activations that process raw (128 + 1) channel EEG signals, progressively downsampling the temporal dimension while extracting multiscale spectral-temporal patterns into 512-dimensional latent feature representations. The second component is the \textbf{BENDRContextualizer}, an 8-layer transformer encoder with 8 attention heads per layer that models long-range temporal dependencies through multi-head self-attention mechanisms, capturing complex contextual relationships in the encoded EEG features across extended time windows. The third component is the \textbf{Regression Head}, consisting of multi-head attention pooling with a learnable CLS token for temporal aggregation, followed by a deep feedforward network (512$\rightarrow$256$\rightarrow$128$\rightarrow$1 neurons) with layer normalization and dropout ($p=0.4$) that maps the contextualized representations to continuous behavioral performance scores. Training proceeds in three phases: Phase 1 trains only the regression head with frozen encoder and contextualizer; Phase 2 unfreezes the contextualizer while keeping the encoder frozen; Phase 3 enables full end-to-end fine-tuning of all components with differential learning rates (head: $5\times10^{-3}$, contextualizer: $5\times10^{-4}$, encoder: $5\times10^{-5}$).

\end{document}